\documentstyle[preprint,aps]{revtex}
\begin{document}
\def\up{\uparrow}
\def\down{\downarrow}
\def\eps{\epsilon}
\draft

\title {
Remarks on Spin Gaps and Neutron Peak Selection Rules in YBCO ---
Do Interlayer Tunneling and Interlayer RVB as Mechanisms for
Cuprate Superconductors Differ?} 

\author{Philip W. Anderson}

\address{Joseph Henry Laboratories of Physics, Princeton
University, Princeton, N.J. 08544 }

\date{March 15, 1996}

\maketitle

\begin{abstract} {We point out that both superexchange between $CuO_2$
layers, and interlayer tunneling, derive from frustrated
one-particle hopping between layers, and that they should be
treated on an equal footing. Doing so, we arrive at a new view of
the nature of pairing in the cuprate supercondcutors, which
explains the striking even $\leftrightarrow$ odd selection rule
observed by Keimer in neutron scattering by YBCO.}
\end{abstract}

\vfill\eject
\narrowtext

Recently, two different hypotheses have been put forward for the
``spin gap'' phenomenon exhibited by the bilayer structures of
cuprate superconductors. (As in YBCO and BISCO 2-layer
materials). Millis and Monien and co-workers\cite{1} have proposed that
what is occurring is pair formation between the two layers with
one of the pair on each layer, motivated by the interlayer
antiferromagnetic superexchange. Strong and myself\cite{2} have
proposed that the interlayer pair tunneling Hamiltonian produces
a correlated state with pairing on each layer for each momentum
$k$, but the different $\vec k$'s are independently phased. Both
postulate ``preformed BCS pair'' states as the essential nature of
the spin gap phenomenon.

Although it is not explicitly mentioned in the former paper, both
theories have one vital element in common: the coherent motion of
single electrons between the two layers must be blocked. As I
have emphasized elsewhere,\cite{3}
antiferromagnetic ``superexchange'' interaction exists only if
the single-particle kinetic energy is frustrated, and is a
consequence of the second-order, virtual action of this kinetic
energy, causing hopping of electrons between the relevant sites.
If the kinetic energy were diagonalized, the Fermi surface would
be split into separate surfaces for even and odd linear
combinations of the two layers, which, being orthogonal, would
exhibit only ferromagnetic exchange interactions. (The reader
should note that although the rest of this article will be couched
in language referring only to bilayer materials, only a slight
modification allows one to present similar arguments for coupled
layers either single or multiple layers.) 

In fact, the Millis-Monien et al papers miss another feature of
interlayer superexchange. 
The hopping matrix element $t_\perp$ between layers
is of course diagonal in $k_{||}$, the interlayer momentum, so
that superexchange, $\sim t^2_{\perp}/U$, must also contain
an extra momentum $\delta$-function. That is,

\begin{eqnarray}
	{\cal H}_{SE}=\sum_k\sum_{k'} c^{+(1)}_{k\up}\
	c^{(2)}_{k\up}\ c^{+(2)}_{k'\down}\ c^{(1)}_{k'\down}
\\\nonumber
	\times t_k\cdot t^*_{k'}/{\rm energy\  denom}
\end{eqnarray}
and the only way to satisfy the pairing condition is for
$k'=-k$. Thus these two mechanisms share the characteristic 
$k$-diagonality, which leads to 
approximate independence of different parts of the Fermi surface;
and the Strong-Anderson mechanism for producing a spin gap is
common to both.

A third relevant theory paper is the weak-coupling solution of the
two-chain Hubbard model by Balents and Fisher\cite{4}, in which
much of the relevant region of parameter space seems to exhibit
the $\rm\underline { same}$ fixed point with a spin gap and one
gapless charged excitation, which seems to interpolate between
the pairing schemes of Ref. (1) and (2).

We argue here that the two mechanisms for spin gaps are not
incompatible but complementary. They result from the same
phenomenon of virtual hopping between layers, which is of course
only virtual because the direct coherent hopping at zero
frequency is blocked by 
interaction effects. Both the spin gap
phenomenon and superconductivity itself are explicable on the
same basis.

We may understand the superexchange phenomenon in Mott insulators
by realizing that the orbitals of opposite-spin electrons need
not be orthogonal. Thus a possible path for exchange of two
electrons of opposite spins is for the second electron (of down
spin, say) to hop back from site 2 to site 1, during the period
while the electron of up spin, while nominally on site 1, has
made a virtual transition to site 2. Thus if $a^+_{1\sigma}$ and
$a^+_{2\sigma}$ are the properly orthogonalized one-electron
operators for the two sites, the actual one-electron operators
are
\begin{displaymath}
	c^+_1=(a^++\epsilon a^+_2)/\sqrt{1+\eps^2}
\end{displaymath}
\begin{displaymath}
	c^+_2=(a^+_2+\epsilon a^+_1)/\sqrt{1+\eps^2}
\end{displaymath}
where
\begin{displaymath}
	\eps=t/U.
\end{displaymath}
These are not orthogonal because they need not be if they belong
to opposite-spin electrons. The 
superexchange energy is
then
\begin{displaymath}
J=\eps t
\end{displaymath}

This picture can be renormalized, as discussed in
Herring\cite{5}, by taking into account interactions between the
two electrons along the exchange path, but the physics remains
the same.

$J$ is the amplitude for interchange of spins keeping the same
charge state. We may also ask for the pair tunneling amplitude,
i.e., the amplitude for an up-spin down-spin pair to tunnel from
one site to another. This is of course charge disfavored for
atomic sites, but for tunneling between metallic layers there is
no charge rigidity. In this case the down-spin electron hops in
the opposite direction, and the amplitude is also
$$E_J=t_\perp\eps\ \ ,$$
with one hop taking place by virtue of non-orthogonality of wave
functions in the two layers, the second by the actual frustrated
one-electron matrix element. This two-particle tunneling process
is the basis of the interlayer theory. It, like superexchange,
 can
be described by an effective Hamiltonian for low energy states.

It becomes clear that there is no sharp way to distinguish
between antiferromagnetic superexchange and pair tunneling
mechanisms for spin gaps and superconductivity if we look at the
order parameters which can result. 

First let us set up some notation. Let $k$ be a momentum near the
Fermi momentum of the two chains or planes, 1 and 2. Orthonormal
electron operators are defined as
\begin{eqnarray}
   a^{(1)+}_{k\up}=a^+_1       &\qquad\qquad
a^{(2)+}_{k\up}=a^+_2\\\nonumber 
   a^{+(1)}_{-k\down}=a^+_{-1} &\qquad\qquad a^{(2)+}_{-k\down}=a^+_{-2}
\end{eqnarray}
and these can combine into the even and odd eigen-operators of
the kinetic energy, 
\begin{eqnarray}a_e^{+}={a^+_1+a^+_2\over\sqrt{2}}\ \  {\rm
etc}.
\\\nonumber 
a_o^{+}={a^+_1-a^+_2\over\sqrt{2}}\ \ {\rm etc}.
\end{eqnarray}
We know that the kinetic energy operator may be written
\begin{eqnarray}
	{\cal H}_k=\eps_k(n_{k_1}+n_{k_2})\\\nonumber
	+ t_\perp (a^+_1\,a_2+a^+_{-1}\,a_{-2})\\\nonumber
	= (\eps_k+t_\perp)\ (a^+_o\,a_o+a^+_{-o}\,a_{-o})\\\nonumber
	+(\eps_k-t_\perp)\ (a^+_e\,a_e+a^+_{-e}\,a_{-e})
\end{eqnarray}
but in the case where interactions within the layers or chains
are sufficiently strong, the splitting given by the last
expression into even and odd eigenstates is not expressed in the
actual state: $t_\perp$ causes primarily virtual
transitions\cite{6}, as in the Mott insulator. (More precisely:
$t_\perp$ causes no $\rm\underline { coherent}$ transitions).
This is an experimental fact in several of the cuprates,
according to photoemission and infrared
spectroscopy.\cite{7,8}.
Instead of splitting, the electrons hop virtually, so that the
electron operators which describe the actual eigen-excitations
are roughly 
\begin{eqnarray}
	c^+_1=(a^+_1+\eps a_2^+)/\sqrt{1+\eps^2}\\\nonumber
	c^+_2=(a^+_2+\eps a_1^+)/\sqrt{1+\eps^2}
\end{eqnarray}

Now let us imagine that by the pair tunneling argument we have
derived an order parameter in which the electrons are paired in
their separate layers:
\begin{eqnarray}
     {\rm (O.P.)}_{\rm TL}
      =
      \langle c^+_1\ c^+_{-1}+c^+_2\ c^+_{-2}\rangle\\\nonumber
      ={
         \langle(a^+_1\ a^+_{-1}+a^+_2\ a^+_{-2})\rangle + \eps
         (a^+_1\ a^+_{-2}+a^+_2\ a^+_{-1})\rangle\over
         \sqrt {1+\eps^2}
       }
\end{eqnarray}
This order parameter is, in terms of orthonormal states, a
mixture of the Strong and the Millis pairings; it may also be
written
\begin{eqnarray*}
{\rm (OP)_{IL}}=
{
   \langle a^+_e\ a^+_{-e}+a^+_o\ a^+_{-o}\rangle\over \sqrt{1+\eps^2}
}\\
+\langle (a^+_e\ a^+_{-e}-a^+_o\ a^+_{-o})
{\eps\over\sqrt{1+\eps^2}}\\\nonumber
=\langle a^+_e\ a^+_{-e}\rangle 
\Big ({1+\eps\over\sqrt {1+\eps^2}}\Big )
+ \langle a^+_o\ a^+_{-o}\rangle \ \
\Big({1-\eps\over\sqrt{1+\eps^2}}\Big )\nonumber
\end{eqnarray*}

Equally, we may imagine a Millis-Monien pairing
\begin{eqnarray}
\langle c^+_1\ c^+_{-2} + c^+_2\ c^+_{-1}\rangle\\\nonumber
= {
\langle (a^+_1\ a^+_{-2}+a^+_2\ a^+_{-1})\rangle + \eps
\langle (a^+_1\ a^+_{-1}+a^+_2\ a^+_{-2})\rangle\over
\sqrt{1+\eps^2}}
\end{eqnarray}

Again, in terms of orthogonal orbitals this is a mixture of the
two pairings. In terms of even and odd
\begin{equation}
{\rm (O.P.)_{MM}}=
\langle a^+_e\ a^+_{-e} 
\Big( 
{1+\eps\over \sqrt{1+\eps^2}}
\Big ) 
-
\langle a^+_o\ a^+_{-o}
\rangle\ 
\Big(
{1-\eps\over \sqrt {1+\eps}}
\Big )
\end{equation}

(Incidentally, neither pairing is of the ``$s,-s$'' type
suggested by Scalapino and Yakovenko). The M-M pairing will
also lead to superconductivity because of the $\eps$ term which
represents interlayer pair tunneling, if there is coupling to
other layers as well as to other momenta, through conventional
short-range interactions.

These two pairings have the interesting feature that they
reinforce each other for the even-even pairing, but have
opposite sign for the odd-odd pairing. We suggest here that the
effective pairing Hamiltonians coming from the two sources have
approximately the same coefficient so that in effect we have only
even-even pairing: the order parameter is 
\begin{displaymath}
{\rm O.P.} \simeq \langle a^+_e\ a^+_{-e}\rangle,
\end{displaymath}
since the odd-odd pairing is favored with $-$ sign by the
superexchange term and favored with $+$ sign by the interlayer
term approximately equally.

This pairing is that appropriate to explain the selection rules
observed by Keimer et al\cite{9} in the scattering of neutrons by the
bilayer material $YBa_2Cu_3O_7$. The observation is that a
pronounced peak appears, rather sharp in energy at $\sim 42$ mev
but broader than instrumental resolution in momentum space, near
$\pi, \pi$ in the $ab$ plane Brillouin zone. The peak appears
only below $T_c$ (magnetic scattering reported above $T_c$ seems
to have been an artifact) and is either simply quasiparticle pair
production, or the same somewhat enhanced by excitonic
interaction effects. More exotic explanations seem incompatible
with the experimental facts, in particular, exotic collective
excitations seem to have no reason to appear sharply below $T_c$,
and at an energy so close to the supposed value of $2\Delta$. The
$k_{||}$-dependence is very compatible with the idea that
coherence factors forbid magnetic scattering between $k$'s with
energy gaps which have the same sign, and enhance strongly those
at energy gaps of opposite sign, which presumably (using BISCO as
a model, and relying on Josephson interference measurements)
occur near points $X$ and $Y$, separated by $\pi,\pi$ in the
zone. The observed BISCO gap would fit the $k_{||}$-dependence
well. 

The dependence on $k_\perp$ in the $c$-direction is remarkable.
This is a sinusoidal curve with a period given by the inverse of
the interplanar spacing, showing that the scattering changes sign
between the two planes. An equivalent statement is that
scattering satisfies the selection rule even $\leftrightarrow$
odd as far as symmetry in the pair of planes is concerned.

A little thought convinces one that this cannot be a coherence
factor selection rule. That is, if the coherence factor is large
between $k$-points $(0,\pi, 0$ and $\pi,0,\pi)$, the gaps
$\Delta_{0\pi 0}=-\Delta_{\pi 0\pi}$. But then the sign of
$\Delta_{\pi 00}$ must be opposite to either one or the other,
leading to a second peak, which is not observed, either at
$\pi,\pi, 0$ or $0,0,\pi$. We propose that this rule holds because
$\rm\underline { at }$ the energy gap the 
even state is preferentially occupied, the
odd state empty.  Then the dominant scattering process for a
neutron is even $\rightarrow$ odd, when the pair of particles
created is primarily at the energy gap. 

The understanding of how this comes about requires us to go
rather
 deeply into the BCS mechanism and the slight generalization that
is necessary in this problem.

In conventional BCS, the effective Hamiltonian for a single pair
of states is
\begin{eqnarray}
	{\cal H}_{\rm one \ pair}
	=(\eps_k-\mu)(n_{k\up}+n_{-k\down}-1)\\\nonumber
	+\Delta_{k}c^+_{k\up}\ c^+_{-k\down}+\Delta^+_k\ c^+_{-k\down}\
	c^+_{k\up}
\end{eqnarray}
An irrelevant constant term has been added to make it clear that
${\cal H}$ may be taken such that it has no effect on the subspace
$n_{k\up}+n_{-k\down}=1$ of states not satisfying the Schrieffer
pairing condition. 

In the pseudospin representation,\cite{10} one may write (1) in
terms of the Nambu spinors $\tau$
as
\begin{equation}
{\cal H}=(\eps_k-\mu)\,\tau^z_k\ +{\Delta\over 2}
\tau^+_k+{\Delta^*\over 2}\tau^-_k
\end{equation}

and it can be diagonalized by 

\begin{equation}
\Psi_o=u_k\ \Psi (\tau^z_k=+1)+v_k\ \Psi(\tau^z_k=-1)
\end{equation}

with

\begin{displaymath}
u_k=\cos{\theta\over 2}, v_k={\sin\theta_k\over 2}
\end{displaymath}

\begin{displaymath}\tan\theta_k={\Delta_k\over\eps_k-\mu}
\end{displaymath}

In the present instance, for a pair of planes we have four
Fermions belonging to a given $k$-vector, $c^{+(\alpha)}_{k\up},
c^{+(\alpha)}_{-k\down}$ with $\alpha=1,2$ a plane index.
Alternatively, we may use $c^{e +}_{k\up}, c^o_{k\up}$ with
\begin{equation}
c^{e,o}_{k\up}={1\over\sqrt {2}}\Big ( c^{(1)}_{k\up}\pm
c^{(2)}_{k\up}\Big ).
\end{equation}

There are two calculations which may be carried out. One is in
the spin-gap situation, where we neglect all coupling to other
momenta. Then the states of momentum $k$, as a decoupled
subspace, couple through the pair tunneling and superexchange
interactions. These are, first, from (1)
\begin{eqnarray*}{\cal H}_{\rm SE}= \sum_k 
{\lambda_k\over 2}\ ( c^{+(1)}_{k\up}\ c^{+(2)}_{-k\down}\
c^{(1)}_{-k\down}\ c^{(2)}_{k\up}\\\nonumber
+(1\leftrightarrow 2) + {\rm H.C.} \nonumber
\end {eqnarray*}
and, second, as previously proposed, 
\begin{displaymath}{\cal H}_{\rm PT}= \sum_k 
{\lambda_k\over 2}\ ( c^{+(1)}_{k\up}\ c^{+(1)}_{-k\down}\
c^{(2)}_{-k\down}\ c^{(2)}_{k\up}
\end{displaymath}
In a similar way to the BCS case, states not satisfying
$n=2$ are annihilated by the interaction. In this case the
Hamiltonian is number-conserving overall and the state which
diagonalizes the sum of the two interactions is simply
\begin{displaymath}\Psi_{sg}=c^{e+}_{k\up}\ c^e_{-k\down}\
\Psi_{\rm vac}
\end{displaymath}
and all other states are unaffected by the pairing Hamiltonian.
This is the spin gap state, identical to that proposed by Strong
and Anderson except that the pairing is in the even state, while
the three other $N=2$ states are either repulsive or not lowered
in energy. Otherwise the story is unchanged. In the
superconducting state we presume the gap function is correlated by
couplings to other nearby momentum states and to other planes. We
can model this by assuming that the momentum-conserving
$\delta$-function is not exact but has a finite width $\delta
k\sim{1\over L}$, (an additional length scale which enters the
interlayer theory, whose physical consequences we do not explore
here). But $\L\to\infty$ leads to physical nonsense when explored
in too great detail. 

Note that any electron excited from a spin gap state must leave
behind a hole in an even state $c_k^{(e)}$ and any $k$ in the spin
gap state can only accept an electron in an odd state
$c^{+(o)}_k$. Thus the spin gap states exactly satisfy the Keimer
selection rule $e\to o$ for scattering of electrons by neutrons.
This reflects the source of the pairing energy in the kinetic
energy term.
\begin{displaymath}
{\rm K.E.} = t_\perp (n_e-n_o).
\end{displaymath}

Also note that although the odd states are nominally unpaired,
nonetheless adding or removing an electron in an odd state from a
spin gapped k-value destroys the pairing criterion $n=2$ and
costs one unit of pairing energy. Thus there is a gap for all
one-electron excitations, even though only the even state is
paired.

Now we consider the more conventional BCS-like theory. Here we
must require phase-coherence of the pairing among all $k$-states,
so that the pairing Hamiltonian is no longer number-conserving,
and may be treated by the usual mean field theory. But when we
transcribe our second-order pairing Hamiltonian into ``even'' and
``odd'' language, it turns out to read

\begin{eqnarray*}
-{\cal H}=\sum_k\lambda_k\sum_{k-k'<L}\Big
(c^{e+}_{k\up}\ c^{e+}_{-k\down}-c^o_{k\up}\
c^o_{-k\down}\Big )\\\nonumber
\times \Big( c^e_{-k'\down}\ c^e_{k'\up}-c^o_{-k'\down}\
c^o_{-k'\up}\Big )\nonumber
\end{eqnarray*}

Thus it is favorable for pairing to occur in even 
$\rm\underline { or }$ odd states
but not in both together. The resulting problem is more
complicated than BCS; in fact, unlike BCS mean field is not quite
adequate to the solution. Basically, the system goes from 4
states empty
\begin{displaymath}
{\rm \bf (a)}\ \ =(n^e_k=n^o_k=o)\ \ {\rm to\ all \ full}\ \ {\rm
\bf(b)}\ \ 
(n^e_k=n^o_k=2)
\end{displaymath}
via 
$\rm\underline { even \ pairing \ only }$ 
\begin{displaymath}
{\rm \bf (c)} \ \ n^e_k=2, n^o_k=o
\end{displaymath}
but continuously, with coherent occupation of (a), (b), (c). A
self-consistent
BCS calculation gives us
\begin{displaymath}
\Delta^{+(k)}_{eff}=\lambda_k\sum_{|k'= k|\langle L} \Big
\langle c^{+e}_k\ c^{+e}_{-k}\rangle - \langle c^{o+}_{k}\
c^{o+}_{-k}\rangle\Big )
\end{displaymath}
changing sign and therefore vanishing at $k=k_F$.\cite{11} What we expect
will happen, however, is that at $k\simeq k_F$ there will be
phase-correlated fluctuations of $b^+_k$ and $b^+_{k'}$ as in the
spin gap state. Thus the actual energy gap does not actually vanish at any $k$.
How to deal in any exact way with the phase fluctuations of the
gap is as yet an unsolved problem. I speculate that the
mathematics of gap formation may resemble more nearly that of
Bose condensation of preformed pairs than the simple pair
condensation of BCS. 

If we had only the same-layer pairing Hamiltonian which we have
used in the past, i.e.,

\begin{displaymath}
\sum_k\lambda_k\sum_{k'-k<L} \Big (c^{+(1)}_k\ 
c^{+(1)}_{-k}\ c^{(2)}_{-k'}\ c^{(2)}_{k'} \ \ +cc\Big ),
\end{displaymath}
we obtain $\lambda_k\chi^{pr}_k\simeq 1$ as the effective gap
equation, with $\chi_k={1\over E_k}$; this can only be satisfied
with
$\epsilon^{2}_k\ \langle \lambda^2_k$. The pairing amplitude
$b_k=\Delta_k=\sin\theta_k$.
\begin{displaymath}
\cos\theta_k={\eps_k\over \lambda_k}=n_k.
\end{displaymath}

A crude approximation is to assume that this same mathematics
occurs twice, once for even pairs and once for odd ones. 
Basically, $n^o_k$ goes linearly from 2 to 0 from
$\eps_k=-\lambda_k$ to 0; and $n^e_k$ goes linearly from 2 to 0
from $\eps_n=0$ to $\eps_k=+\lambda_k$.

The ratio of amplitudes $e\to e$ and $o\to o$ to
$e\leftrightarrow o$ can be
calculated using this assumption naively. The ratio
of occupation factors is equal to 
\begin{eqnarray*}
{
\int^\lambda_{-\lambda}n^o_k d\eps\int^\lambda_{-\lambda}
d\eps(2-n^e_k)+\int^\lambda_{-\lambda}n^e_k
d\eps\int^\lambda_{-\lambda}n^o_k)d\eps
\over
\int^\lambda_{-\lambda}n^e_k d\eps\int^\lambda_{-\lambda}
(2-n^e_k)d\eps+\int^\lambda_{-\lambda}n^o_k\
\int^\lambda_{\-lambda} (2-n^o_k)
}={5\over 3}
\end{eqnarray*}

This leaves out coherence factors which may considerably enhance
the ratio (I estimate by a factor 2). 
This is still less skewed than the data. 
The state may resemble 
more the correlated
``spin gap'' state than this uncorrelated mean field
approximation. The ``spin gap'' give a ratio of 1:0, and it is
unreasonable that the superconducting state should be very much lower. 

In a forthcoming paper, in collaboration with S. Chakravarty, we
shall show how the amplitude, shape and intensity of the neutron
peak follows from the above ideas. Many new complexities are
caused by the unique nature of the pairing Hamiltonian in this
theory, and we look forward to exploring an unexpectedly rich
field of physics.

I have benefitted greatly from discussions with N.P. Ong, D.C.
Clark, S.P. Strong and, especially, S. Chakravarty.
The work grew out of discussions with B. Keimer. 
This work was supported by the NSF, Grant \# DMR-9104873.

\end{document}